%% file: gorelov-had2011.tex
\documentclass[a4paper,11pt]{article}

\usepackage{longtable}
\usepackage{hyperref}
\usepackage{contribution}
\usepackage{subfigure}

\input{econfmacros}
\input{contributionmacros}
\input{symblibrary.tex}

\begin{document}

\input{contribution}

\end{document}

%% file: econfmacros.tex
\newcommand{\weblink}[2][]{%
    \ifthenelse{\equal{#1}{}}%
    {\textnormal{\url{#2}}}%
    {\textnormal{\href{#2}{#1}}}%
}

\newcommand{\acknowledgements}[1]{%
  \bigskip\bigskip
  \textsf{\textbf{\Large Acknowledgements}} \\[2ex]
  {#1}
  \bigskip
}

\def\beq{\begin{equation}}
\def\eeq#1{\label{#1}\end{equation}}
\def\eeqn{\end{equation}}

\def\beqa{\begin{eqnarray}}
\def\eeqa#1{\label{#1}\end{eqnarray}}
\def\eeqan{\end{eqnarray}}

\let\bar=\overbar

\def\D{{\cal D}}

\def\O{{\cal O}}

\def\Dslash{\not{\hbox{\kern-4pt $D$}}}
\def\dslash{\not{\hbox{\kern-2pt $\del$}}}

\def\BR{\mbox{\rm BR}}

\def\msb{{\bar{\ssstyle M \kern -1pt S}}}

\def\s#1{\widetilde{#1}}

%% file: contributionmacros.tex
\newcommand{\contribution}[7][]{%
  \clearpage
  \thispagestyle{plain}
  \ifthenelse{\equal{#1}{}}
  {\hypersetup{pdftitle={#2}}}
  {\hypersetup{pdftitle={#1}}}
  \hypersetup{pdfauthor={{#3} {#4}}}
  {\centering\normalfont\LARGE\bfseries\sffamily #2 \par\nobreak}
  \lhead{}
  \chead{%
    \textit{\footnotesize XIV International Conference on Hadron Spectroscopy
      (\weblink[\textit{Hadron2011}]{http://www.hadron2011.de}), 13-17 June 2011, Munich, Germany}%
  }
  \rhead{}
  \bigskip
  \begin{center}
    {#3} {#4}\ifthenelse{\equal{#6}{}}{}{\footnote{\weblink[#6]{mailto:#6}}}
    \ifthenelse{\equal{#7}{}}{}{#7} \\
    \textit{#5}
  \end{center}
  \bigskip
}

\renewcommand{\abstract}[1]{%
  \begin{center}
    \begin{minipage}{0.85\textwidth}
      \begin{footnotesize}
        #1
      \end{footnotesize}
    \end{minipage}
  \end{center}
  \bigskip
}

%% file: contribution.tex
%
%
%
%
%
{  


%

\contribution[Heavy Hadrons at the Tevatron]  
{Heavy Hadron Spectroscopy and Production at the Tevatron}  
{Igor V.}{Gorelov}  
{Department of Physics and Astronomy, \\
  University of New Mexico, MSC07 4220,\\
  1919 Lomas Blvd. NE, \\
  Albuquerque, NM 87131-0001, USA}  
{gorelov@fnal.gov}  
{on behalf of the CDF and \( \dnull \) Collaborations}  
%

%
\abstract{%
           Using data from \(\ppbar \) collisions at
           \(\sqrt{s}=1.96\tev\) recorded by the \cdf2 and \( \dnull \)
           detectors at the Fermilab Tevatron, we present recent
           results on charm and bottom hadrons. We the most recent
           CDF results on properties of the four bottom baryon resonant
           states \(\Sgbstm,\,\Sgbstp\).  New results on exotic
           \(Y(4140) \) state observed by CDF are also reported.  A precise
           measurement of production rates of the lowest lying
           bottom baryon, \(\Lb \), produced in the \( \dnull \) detector
           is presented.  
         }%
%
%

%
\section{
  Measurement of the Masses and Widths of the Bottom Baryons
  \( \mathbf{\Sgbstm} \) and \( \mathbf{\Sgbstp} \) with the {\bf CDF~II} Detector
}
  Baryons with a heavy quark \( Q \) as the ``nucleus'' and a light
  diquark \( q_{1}q_{2} \) as the two orbiting ``electrons'' are the
  helium atoms of QCD.  The heavy quark in the baryon may be used as a
  probe of confinement which allows the study of non-perturbative QCD in
  a different regime from that of the light baryons.
\par 
  A recent comprehensive review of the experimental and theoretical
  status of baryon spectroscopy with many useful references can
  be found in~Ref.~\cite{Klempt:2009pi}. The resonant \(\Sgbst\) states
  have been discovered by CDF~\cite{:2007rw}, and this study follows that
  first observation.
\par 
  The \(\Sgbstpm \) candidates are reconstructed in their exclusive
  decay modes to \( \Lb\pipm_{\mathit{soft}} \). The base state \Lb is
  reconstructed in its weak decay \( \Lb\to\Lc\pim_{b} \) with the
  \(\Lc\) candidates found in the \(\LcpKpi\) decay by fitting three
  tracks to a common vertex.  The \Lb vertex is formed by a \(\Lc\)
  candidate combined with a fourth pion track, the \(\pim_{b}\), having
  a transverse momentum above \(1.5\gevc \). Then the vertex is
  subjected to a three-dimensional kinematic fit.  The \Lb signal in the
  invariant mass distribution \(M(\Lc\pim_{b})\) amounts to approximately
  \( 16\,300 \) candidates at the expected \Lb mass, with a signal to
  background ratio around \(1.8\).
  To produce the spectra of \(\Sgbstpm\to\Lb\pipm_{\mathit{soft}}\)
  candidates, each \(\Lc\pim_{b}\) candidate from the \Lb signal region
  of \(m(\Lb)\in(5.561,\,5.677)\gevcc\) is combined with one of the
  tracks remaining in the event with transverse momentum above
  \(200\mevcc \) and with a pion mass hypothesis assigned.  The analysis
  of the \Sgbstpm signals is performed using the mass difference
  distributions
    \(Q = m(\Lb\pipm_{\mathit{soft}}) - m(\Lb) - m(\pipm)\), 
  where \( m(\pipm) \) is set to its world-average
  value~\cite{Nakamura:2010zzi}.  The mass resolution of the
  \(m(\Lb\to\Lc\pim_{b})\) signal and most of the systematic
  uncertainties cancel in the mass difference spectrum, yielding fine
  detector resolution in the \(Q\)-value scale.
  The experimental \Sgbstm and \Sgbstp \(Q\)-value distributions, each
  fitted with an individual unbinned maximum likelihood (ML) functions, are
  shown in Fig.~\ref{fig:signal-sgbm}.  The projection of the
  corresponding unbinned ML fit is superimposed on each graph.  The
  shape of the signal is modeled with a non-relativistic Breit-Wigner
  function convoluted with a Gaussian detector resolution. As the soft
  pion \(\pipm_{\mathit{soft}}\) is emitted in a \(P \)-wave, the
  Breit-Wigner width parameter is modified by the \(P \)-wave
  factor~\cite{Jackson:1964zd}
\( \Gamma=
   \Gamma_{0}\cdot
   {\left( {p^{*}_{\pi_{\mathit{soft}}}}/{p^{*0}_{\pi_{\mathit{soft}}}} \right)}^{3}
\). 
  The background is described by an
  ordinary second order polynomial modulated by a kinematically motivated threshold factor, 
  specifically \(\sqrt{(Q+m_{\pi})^{2}-{m_{T}}^{2} }\cdot\mathcal{P}^{2}(Q;C,b_{1},b_{2})\).
  The left and right plots in Fig.~\ref{fig:signal-sgbm} show clear signals of 
  \(\Sigbm,\,\Sigbstm \) and \( \Sigbp,\,\Sigbstp \), respectively.
  The significance of every peak is well above \(6\cdot\sigma\). 
\begin{figure}[hbtp]
\begin{center} 
  \includegraphics[width=0.49\textwidth]
  {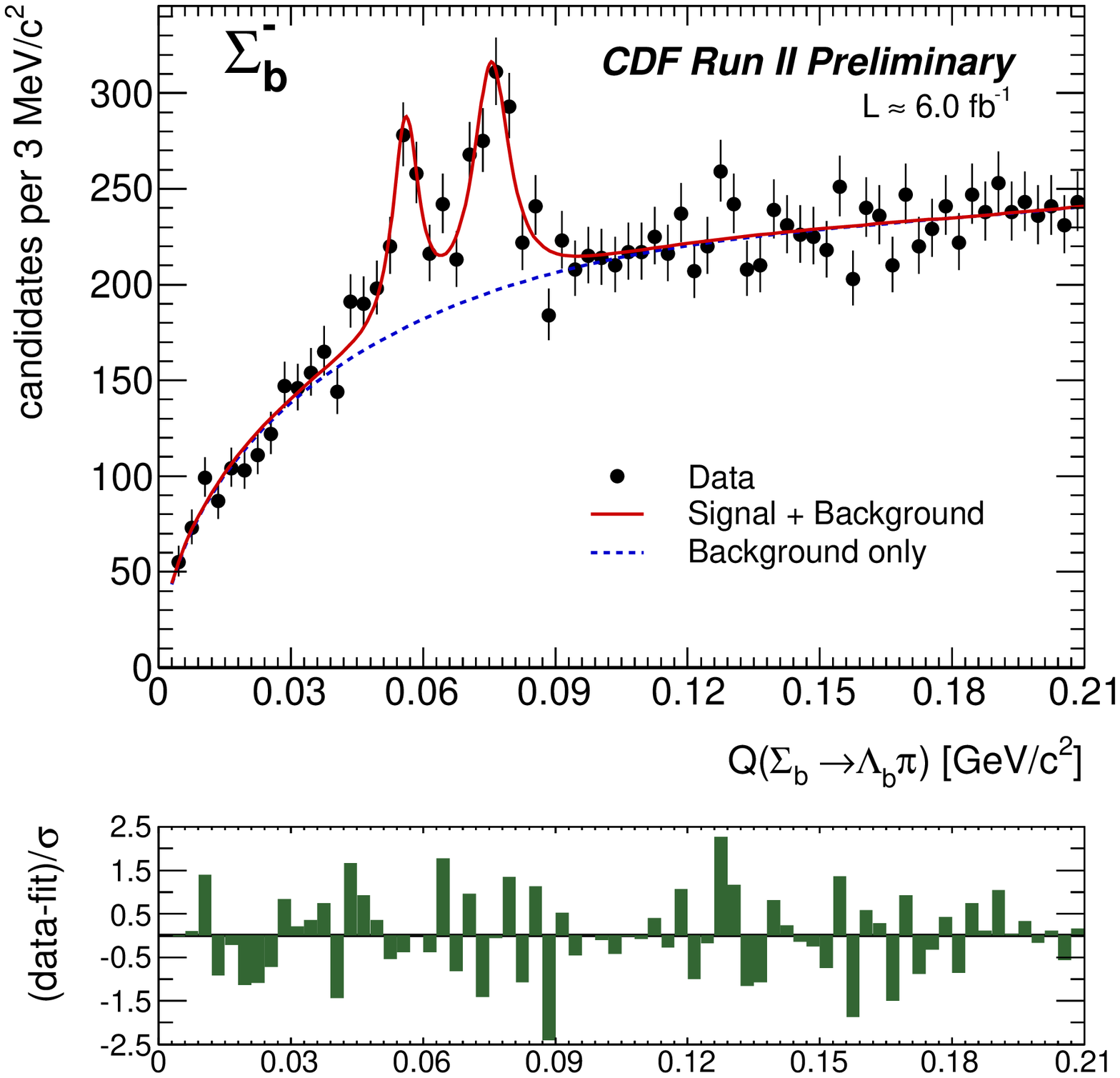} 
  \includegraphics[width=0.49\textwidth]
  {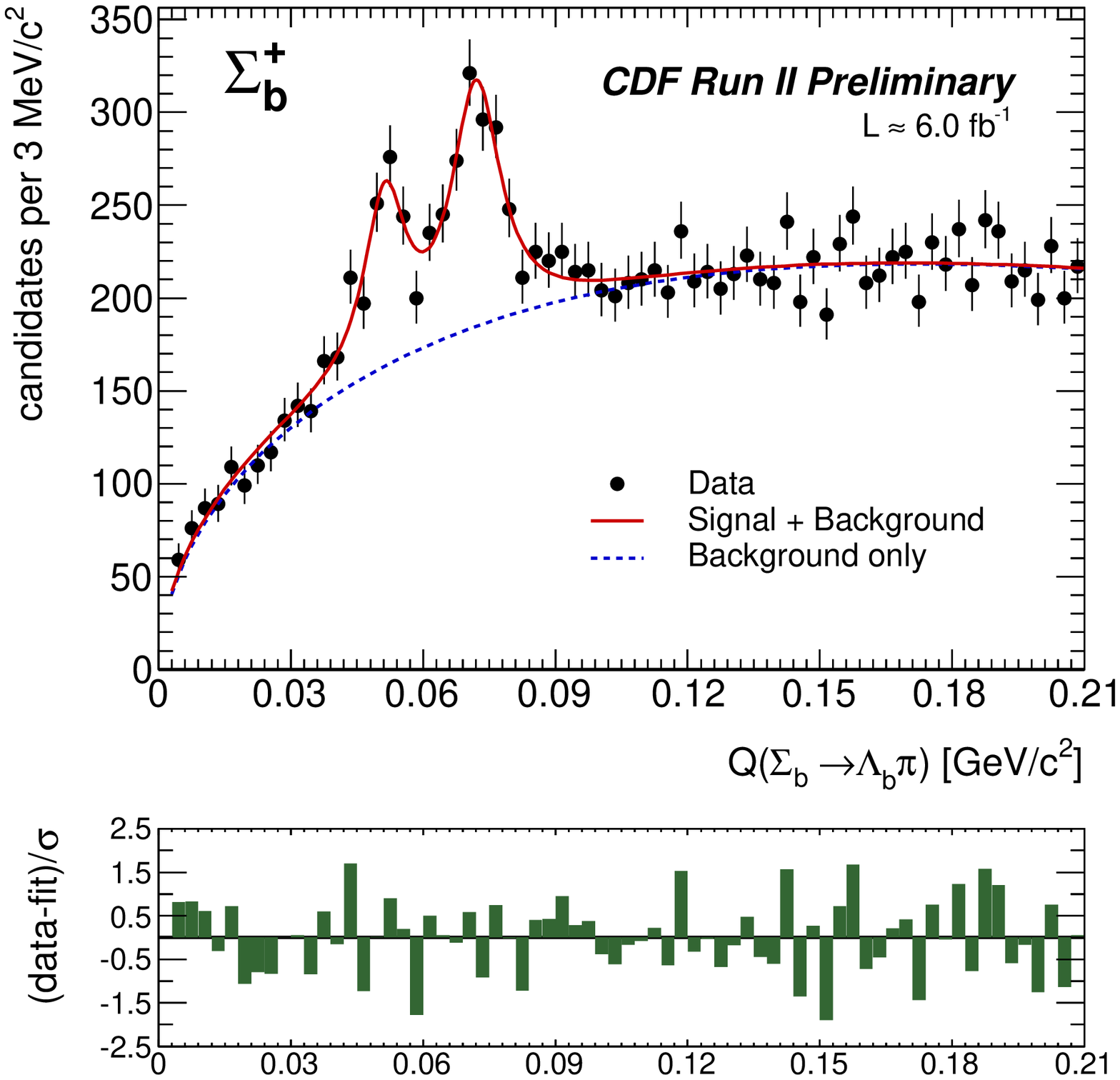}  
  \caption{ The \( Q \)-value spectra for \Sgbstm (left plot) and
            \Sgbstp (right plot) candidates, 
            where \({Q = M(\Lb\pipm)-M(\Lb)-m_{\pipm}}\), 
            are shown with the projection of the corresponding unbinned
            ML fit superimposed on the binned distribution. The
            pull distributions of the fit (bottom plots) are evenly
            distributed around zero with fluctuations of about
            a \(\pm2\sigma\) range. }
\label{fig:signal-sgbm}
\end{center}
\end{figure}
\begin{table}[tb]
\begin{center}
\begin{tabular}{lccc}
\hline
\hline
State        & \(Q\)-value,\mevcc & Absolute mass \(m\),\mevcc  & Width \(\Gamma\),\mevcc \\
\hline
{\Sigbm} & {\(56.2\,_{-0.5\,-0.4}^{+0.6\,+0.1} \)} 
                          & {\(5815.5\,_{-0.5}^{+0.6}\pm1.7 \)} 
                          & {\(4.3\,_{-2.1-1.1}^{+3.1+1.0} \)} \\
{\Sigbstm} & {\(75.7\,\pm0.6\,_{-0.6}^{+0.1} \)} 
                          & {\(5835.0\,\pm0.6\,\pm1.8 \)} 
                          & {\(6.4\,_{-1.8\,-1.1}^{+2.2\,+0.7} \)} \\
{\Sigbp} & {\(52.0\,_{-0.8\,-0.4}^{+0.9\,+0.1} \)} 
                          & {\(5811.2\,_{-0.8}^{+0.9}\pm1.7 \)} 
                          & {\(9.2\,_{-2.9\,-1.1}^{+3.8\,+1.0} \)} \\
{\Sigbstp} & {\(72.7\,\pm0.7\,_{-0.6}^{+0.1} \)} 
                          & {\(5832.0\,\pm0.7\,\pm{1.8}  \)} 
                          & {\(10.4\,_{-2.2\,-1.2}^{+2.7\,+0.8} \)} \\
\hline
%
  & \multicolumn{3}{c}{Isospin mass splitting, \mevcc} \\
\hline
%
 {\(m(\Sigbp) - m(\Sigbm)\)} 
              & \multicolumn{3}{c}{ {\( -4.2\,_{-0.9}^{+1.1}\pm{0.1} \)} } \\
 {\(m(\Sigbstp) - m(\Sigbstm)\)} 
                & \multicolumn{3}{c}{ {\(  -3.0\pm0.9\pm{0.1} \)} } \\
\hline
\hline
\end{tabular}
\caption{ Summary of the final results.  For all the entries, the
         first uncertainty is the statistical one and the second is
         systematic.  To extract the absolute masses, the best CDF mass
         measurement for \Lb~\cite{Acosta:2005mq} has been used. }
\label{tab:results}
\end{center}
\end{table}
  The analysis results are arranged in Table~\ref{tab:results}. 
\par
  In conclusion, the first observation~\cite{:2007rw} of the
  \(\Sgbstpm\) bottom baryons has been confirmed.  The direct mass
  difference measurements have statistical precision a factor of
  \(\gsim2.3\) better than was previously reported~\cite{:2007rw} due to
  the larger dataset.  The measurements are in good agreement with the
  previous results.  The isospin mass splittings within the \( I=1 \)
  triplets of \( \Sigb \) and \( \Sigbst \) states have been extracted
  for the first time.  The \(\Sgbstm\) states have higher mass values
  than their \(\Sgbstp\) partners following a pattern~\cite{Guo:2008ns}
  common to most of the known isospin multiplets. These
  measurements favor the phenomenological explanation of this ordering
  due to higher masses of \(d \)-quarks than \(u \)-quarks
  and a larger electromagnetic contribution due to electrostatic Coulomb
  forces between quarks in \(\Sgbstm\) states than in the \(\Sgbstp\)
  ones.  The natural widths of the \( \Sigbpm \) and \( \Sigbstpm \)
  states have been measured for the first time. The measurements are in
  agreement with theoretical expectations~\cite{Korner:1994nh}, within
  their experimental uncertainties. For further details on this
  analysis, please see~\cite{Calancha:2010dq}.
%
%
\section{
  Update on the \( \mathbf{Y(4140)} \)  Near-Threshold Structure 
  in \( \mathbf{J/{\Psi}}{\phi} \) Mass Spectrum  
  of the \( \mathbf{\Bu\to{J/{\Psi}}} {\phi}\mathbf{\Kp}\) Decays 
}
  Recently, evidence has been reported by the CDF Collaboration for a
  narrow structure near the \( \jpsi\phi \) threshold in exclusive
  \(\Bu\to\jpsi\phi\Kp\) decays produced in \(\pap\) collisions at
  \(\sqrt{s}=1.96~\tev\)~\cite{Aaltonen:2009tz}.  A latest
  update~\cite{Aaltonen:2011at} reports a preliminary confirmation of
  the signal in \( M(\jpsi\phi) \) spectrum. The analysis is
  based on a larger sample of data comprising an integrated luminosity
  of \(\sim6\invpb \) accumulated by the \cdf2 detector. The specific data
  sample is collected by a dedicated three-level dimuon trigger
  recording \( \jpsi\to\mumu \) events.
\par 
  The \(\Bu\to\jpsi\phi\Kp\) candidates are reconstructed from
  \(\jpsi\to\mumu \) candidates taken within \(\pm50\mevcc \) around
  the mass of the \(\jpsi \) and \(\phi\to\Km\Kp \) candidates within a \(\pm7\mevcc\)
   window at the \(\phi\) mass combined further with the \(\Kp \)
  positive track. All five tracks are subjected to a \(3D \) kinematic
  fit to a common vertex with \(M(\jpsi) \) constrained to its PDG
  value. \(\dedx\) and {\rm ToF} measurements are used to identify three
  kaons contributing to the final state to further suppress the
  combinatorial background.  The total transverse momentum \pt of the final candidate is
  required to be above \(4\gevc \).
  The plot in Fig.~\ref{fig:y4140-a} shows a prominent \(\Bu\)
  meson signal of \( 115\pm12\,\stat\) events in the invariant mass
  spectrum \(M(\jpsi\phi\Kp)\).
\begin{figure}[hbtp]
\begin{center} 
  \subfigure[]{\label{fig:y4140-a}\includegraphics[width=0.43\textwidth]{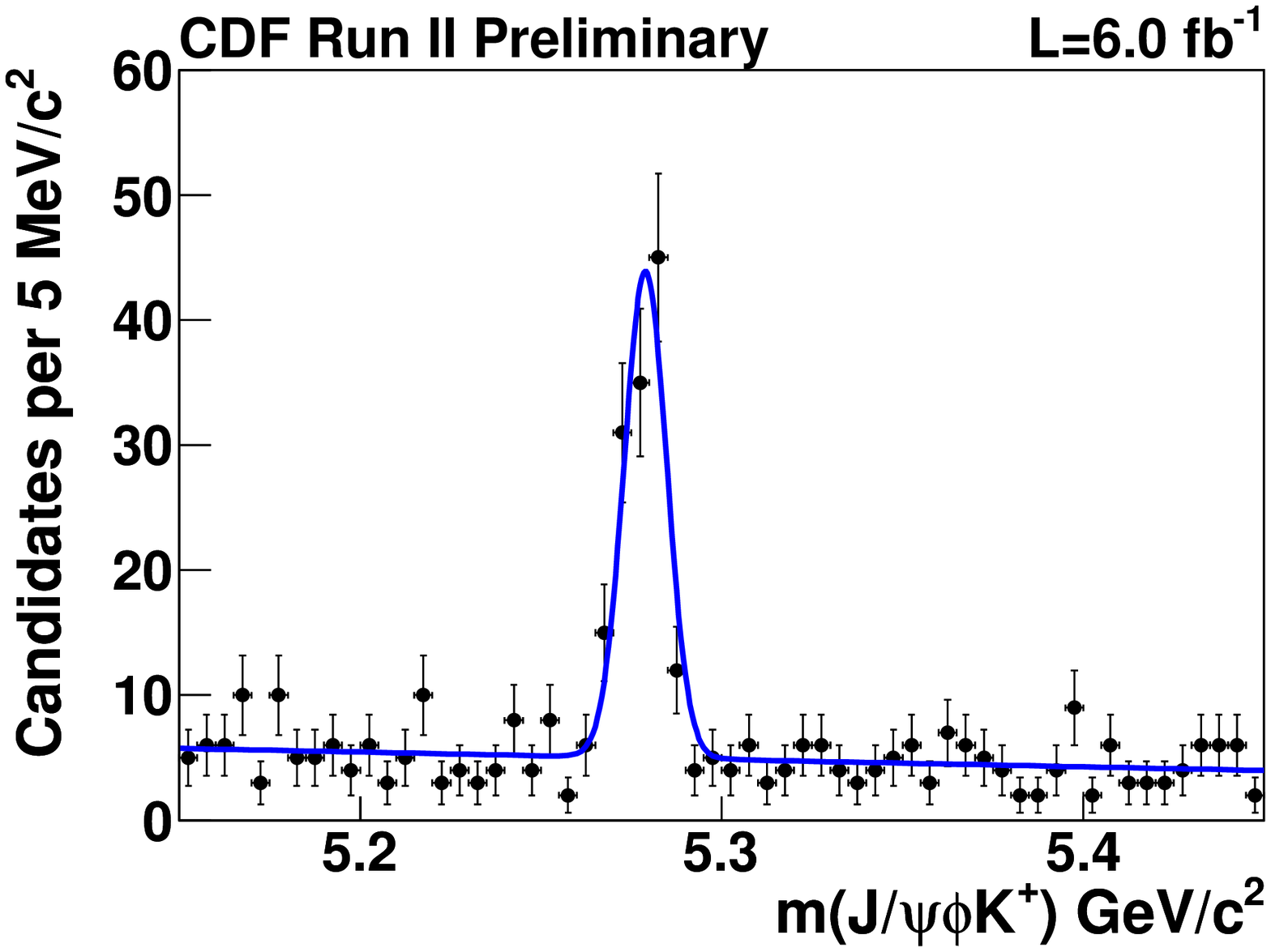}} 
  \subfigure[]{\label{fig:y4140-b}\includegraphics[width=0.43\textwidth]{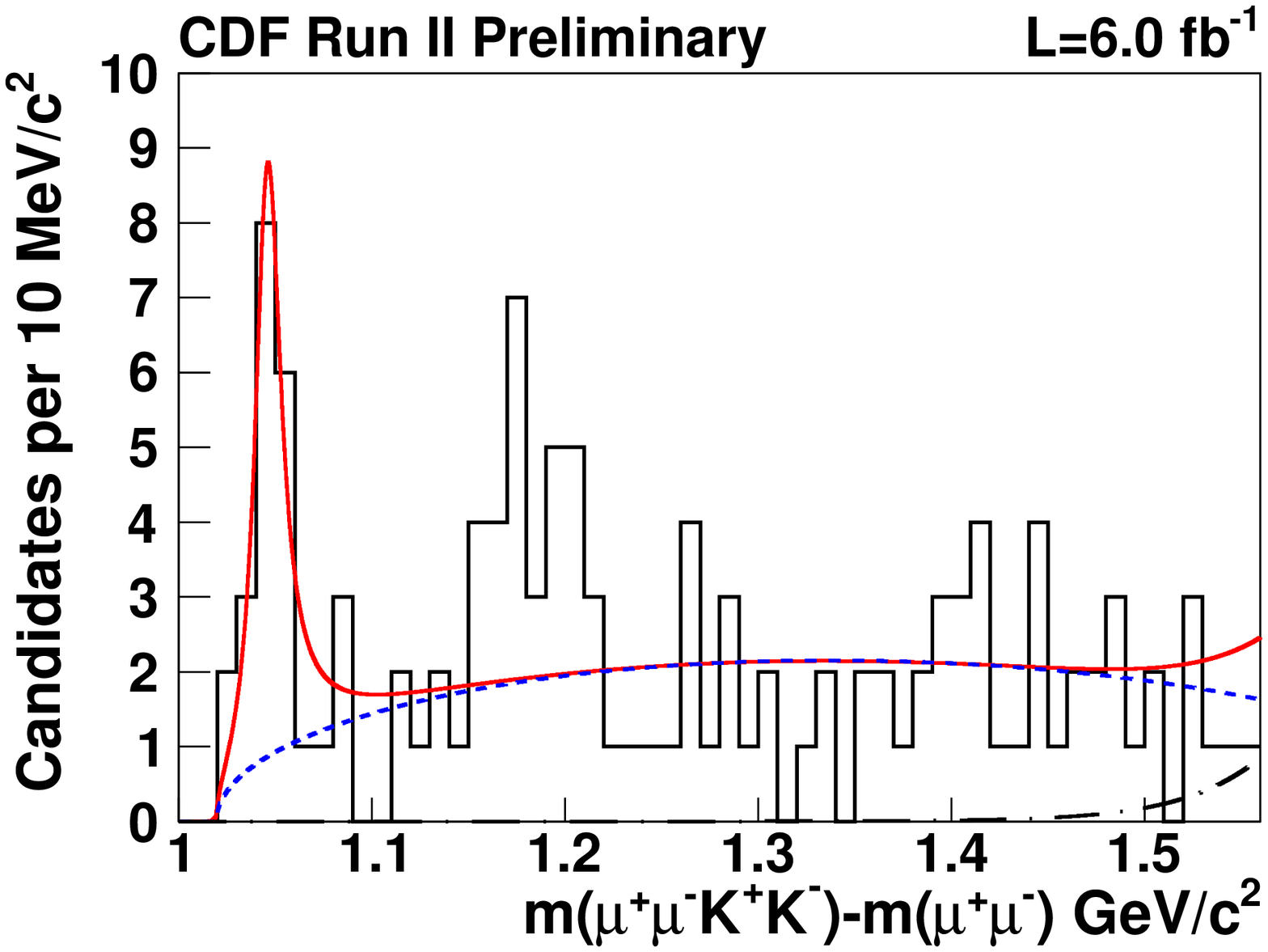}} 
  \caption{
           (a): The \(\Bu\) meson mass distribution
           \(M(\jpsi\phi\Kp)\) is shown with a fit to the data made with
           a Gaussian signal function and a linear background function.
           (b): The mass difference spectrum \( \Delta{M} =
           M(\mumu\Km\Kp) - M(\mumu) \) with the \(\mumu\Km\Kp \)
           combinations contained within \(\Bu\to\jpsi\phi\Kp\)
           candidates at \(\pm3\sigma \) around nominal \( M(\Bu)
           \)~\cite{Nakamura:2010zzi}; the background is predicted by
           the sum of the pure three-body phase space background
           contribution (dotted curve) and the \(\Bs\) meson
           contamination (fixed to \( 3.3\), dash-dotted curve); the
           solid red curve is the total unbinned ML fit where the signal
           PDF is an \(S\)-wave Breit-Wigner convoluted with the
           resolution of \(\sigma=1.7\mevcc \).
}
\label{fig:y4140}
\end{center}
\end{figure}
  Selecting the \(\Bu\) candidates in \(\pm3\sigma \) mass range around
  its nominal mass the mass difference spectrum
    \( \Delta{M} = M(\mumu\Km\Kp) - M(\mumu) \) 
  within final states of those \(\Bu\) candidates is examined. An
  enhancement near a threshold is seen at the 
  Fig.~\ref{fig:y4140-b}. The unbinned ML fit shown at the plot returns
  the signal position, width and its yield arranged in a
  Table~\ref{tab:y4140}.  The fitted parameters are consistent with the
  previous results~\cite{Aaltonen:2009tz} on the \(Y(4140) \) state.
  The \(p\)-value of the observed
  signal with respect to the background is determined using a statistical
  trials generation.  It has been found to be \(p\approx{2.3}\cdot{10^{-7}} \)
  or \(\approx{5.0}\cdot\sigma \) in Gaussian terms.
\begin{table}[tb]
\begin{center}
\begin{tabular}{llllll}
\hline
\hline
  State & \(\Delta{M}\), \mevcc & \(M\), \mevcc & \(\Gamma\), \mevcc & \(N_{\rm cands}\) & \(N_{\sigma}\) \\
\hline
  \( Y(4140) \) & {\({1046.7}^{+2.9}_{-3.0}\)} & {\({4143.4}^{+2.9}_{-3.0}\pm{0.6}\)} 
                & {\({15.3}^{+10.4}_{-6.1}\pm{2.5}\)}
                & {\({19}\pm{6} \)} & \({\approx{5.0}\cdot\sigma }\)\\
\hline
                \multicolumn{6}{l} {
  \(\BR(\Bu\to{Y(4140)}\Kp)\cdot\BR(Y(4140)\to\jpsi\phi)/\BR(\Bu\to{\jpsi\phi}\Kp)=0.149\pm{0.039}\pm{0.024}\)
                                     }\\
\hline
\hline
\end{tabular}
\caption{ Summary of the results found for the \( Y(4140) \). 
          On all the entries, the first uncertainty is the statistical
          one and the second is systematic.  The absolute masses are
          extracted from the fitted \( \Delta{M} \) and the world
          average \(M(\jpsi)\)~\cite{Nakamura:2010zzi}. The measured
          relative branching ratio for \( Y(4140) \) is presented in
          row 2. }
\label{tab:y4140}
\end{center} 
\end{table}
  In conclusion, the increased \(\Bu\to\jpsi\phi\Kp\) sample at CDF
  allows further investigation of the \( Y(4140) \) structure and a
  preliminary observation of the signal is
  reported~\cite{Aaltonen:2011at}. The mass and width are found to be
  consistent with the previous report~\cite{Aaltonen:2009tz}.
\par
  Other experimental
  groups~\cite{Shen:2009vs}~\cite{LHCb:none-4140}~\cite{Stone:2011vd} do
  not confirm the observation of the \( Y(4140) \) structure.  Further
  investigation is needed and work is ongoing at CDF Collaboration to
  update the whole analysis on the full Run II data sample.
%
\section{
  Measurement of the \( \mathbf{\Lb} \) Production Fraction \\
  \( \mathbf{ f(b\to\Lb)\cdot\BR(\Lb\to{J/{\Psi}}\Lz) } \) with the \( \dnull \) Detector
}
  For the \(\Lb \), the lightest bottom baryon, only a few decay channels
  have been studied, and the uncertainties on its branching fractions
  are large, \(\sim(30-60)\%\).
  Increasingly precise measurements of
  \({f(b\to\Lb)\cdot\BR(\Lb\to{\jpsi}\Lz)} \) (where
  \(f(b\to\Lambda_{b})\) is the fraction of \b quarks which hadronize to
  \Lb baryons) will allow better tests of models including PQCD and 
  relativistic and nonrelativistic quark models which predict heavy
  baryon decays. Moreover, these measurements could help in the study of
  \(\b\to\s\) transitions such as
  \(\Lb\to\mumu\Lz\)~\cite{Lbrare1}~\cite{Lbrare2}, which are topologically
  similar to \( {\Lb\to{\jpsi}\Lz}\) where \( {\jpsi\to\mumu} \).
\par
  \( \dnull \) reports an improved measurement (relative to 
  Ref.~\cite{CDFBrLbrecent}) on the relative
  production fraction, specifically
   \[ \sigma_{\rm rel} = \frac{f(b\to\Lb)\cdot\BR(\Lb\to{\jpsi}\Lz)}{f(b\to\Bd)\cdot\BR(\Bd\to\jpsi\KS)}
                       = \frac{N(\Lb)}{N(\Bd)}\cdot
                         \frac{\BR(\Lb\to{\jpsi}\Lz)}{\BR(\Bd\to\jpsi\KS)}\cdot
                         \frac{\BR(\KS\to\pipi)}{\BR(\Lz\to\proton\pim)}\cdot{\epsilon}\,,\]  

  where \(\epsilon=2.37\pm0.05~\mbox{(MC stat.)} \) is the relative
  detection efficiency of the well-measured \(\Bd\to\jpsi\KS\)
   reference signal in denominator with respect to the \(\Lb\to{\jpsi}\Lz\) in
  the numerator.  From this measurement one can extract 
  \({f(b\to\Lb)\cdot\BR(\Lb\to{\jpsi}\Lz)} \) with a significantly
  improved precision compared to the current world
  average~\cite{Nakamura:2010zzi}.  The study is based on
  \(\IntL\approx6.1\invfb \) of \ppbar collisions collected with the
  \(\dnull \) detector between 2002 and 2009.
  The invariant mass distributions of the \(\Lb\) and \(\Bd \)
  candidates passing the analysis criteria are shown in
  Fig.~\ref{fig:lbfrac-a} and Fig.~\ref{fig:lbfrac-b} correspondingly.
  To extract the yields of the observed \(\Lb\) and \(\Bd \) hadrons, an
  unbinned ML fit is applied to each mass distribution assuming a
  double Gaussian function for each signal and a second order polynomial
  distribution for their backgrounds. The fits yield 
  \(N({\Lb\to{\jpsi}\Lz})\,=314\pm29\,{\rm events}\) and 
  \(N({\Bd\to\jpsi\KS})\,=2335\pm73\,{\rm events}\).
\begin{figure}[hbtp]
\begin{center} 
  \subfigure[]{\label{fig:lbfrac-a}\includegraphics[width=0.49\textwidth]{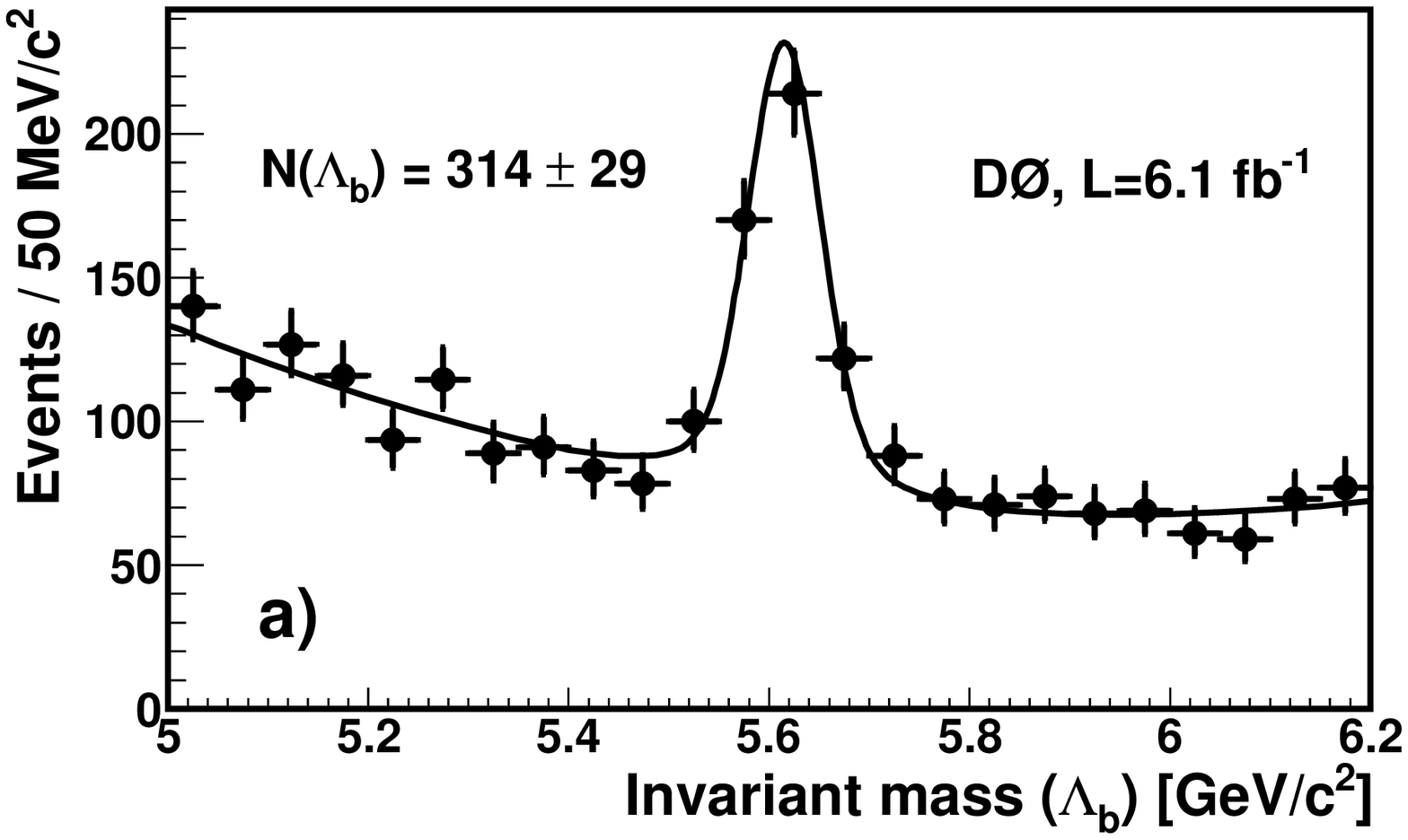}} 
  \subfigure[]{\label{fig:lbfrac-b}\includegraphics[width=0.49\textwidth]{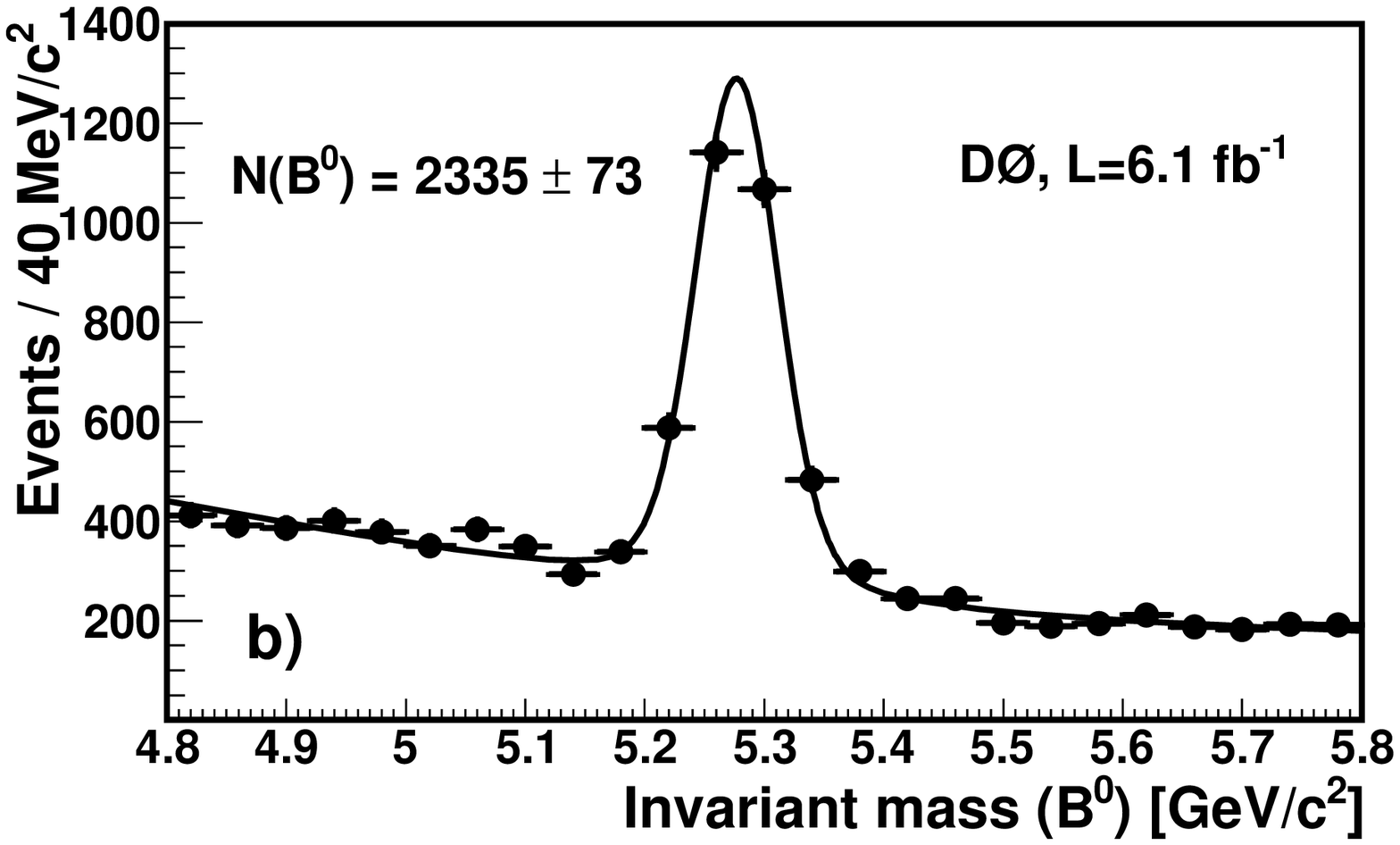}} 
  \caption{ Invariant mass distribution in data for (a) 
            \({\Lb\to{\jpsi}\Lz}\) and (b) \(\Bd\to\jpsi\KS\) 
            decays. Fit results are superimposed.
          }
  \label{fig:lbfrac}
\end{center}
\end{figure}
%
  In summary, the \dnull Collaboration has obtained the production
  fraction multiplied by the branching fraction for the decay
  \({\Lb\to{\jpsi}\Lz}\) relative to that for the decay
  \(\Bd\to\jpsi\KS\) to be
  \[\sigma_{\rm rel}={0.345}\pm{0.034}\stat\pm{0.033}\syst\pm{0.003}\,({\rm PDG})\,.\]
  The measurement is the most precise to date and exceeds the precision
  of the current value reported as the world average, 
  \(0.27\pm0.13\)~\cite{Nakamura:2010zzi}.
  Using the PDG value~\cite{Nakamura:2010zzi},\\
  \({f(b\to\Bd)\cdot\BR(\Bd\to\jpsi\KS)}=(1.74\pm 0.08)\times{10}^{-4}\),
  one extracts
  \[ {f(b\to\Lb)\cdot\BR(\Lb\to{\jpsi}\Lz)} = 
     \left [ {6.01\pm0.60\stat}{\pm0.58\syst}{\pm0.28\,({\rm PDG})} \right ] \times{{10}^{-5}} \]
  which can be compared directly to the world average value of
  \(({4.7}\pm{2.3})\times{{10}^{-5}}\)~\cite{Nakamura:2010zzi}.  
  This result represents a reduction by a factor of \(\sim{3}\) of the
  uncertainty with respect to the previous
  measurement~\cite{CDFBrLbrecent}.  For further details on this
   analysis, please see~\cite{Abazov:2011wt}.
%
%
%
\par
\acknowledgements{
  The author is grateful to his colleagues from the CDF and \(\dnull\)
  {\it B}-Physics Working Groups for useful suggestions and comments
  made during preparation of this talk.  The author thanks the US
  Department of Energy for support of this work.
}


%

}  
